 \documentstyle[12pt]{article}
\newcommand{\be}{\begin{equation}}
\newcommand{\ee}{\end{equation}}
\newcommand{\bea}{\begin{eqnarray}}
\newcommand{\eea}{\end{eqnarray}}

\begin{document}
 \begin{titlepage}

\begin{flushright}
CERN-TH.7021/93
\end{flushright}
\vspace{20 mm}

\begin{center}
{\huge The Deformed Matrix Model}

\vspace{5mm}

{\huge at Finite Radius}

\vspace{5mm}

{\huge and a New Duality Symmetry}
\end{center}

\vspace{10 mm}

\begin{center}
Ulf H. Danielsson\\
Theory Division, CERN, CH-1211 Geneva 23, Switzerland

\end{center}

\vspace{2cm}

\begin{center}
{\large Abstract}
\end{center}
The $1/x^{2}$ deformed $c=1$ matrix model is studied at finite radius
and non-zero cosmological constant. Calculational techniques are
presented and illustrated in some examples. Furthermore, a new
kind of $R \rightarrow 1/R$ duality is discovered, which mixes
different genera.

\vspace{2cm}
\begin{flushleft}
CERN-TH.7021/93 \\
September 1993
\end{flushleft}
\end{titlepage}
\newpage

\section{INTRODUCTION}
Recently
the $1/x^{2}$ deformed $c=1$ matrix model was introduced
by Jevicki and Yoneya in \cite{jev}.
The motivation was to find a matrix-model
description of the eternal two-dimensional stringy black hole
described in \cite{wit}. The new matrix model has been studied in
several subsequent papers, [3--6].
It has been found that the model is clearly different from the
standard $c=1$. Furthermore, it has a few important
properties, which makes it an interesting black-hole candidate,
in particular the scaling properties and the position of poles in
correlation functions.

In this paper the $1/x^2$ deformed $c=1$ matrix model will be
studied at finite radius $R$. This is important in view of the
possible black-hole connection. In this case we would expect a
compactified time coordinate in the Euclidean version.
Unfortunately, the matrix model does not seem to prefer any
particular value of the radius.
Since the black-hole connection is just an educated guess, and
we really have no good understanding of the physics of the deformed
model, it is important to learn as much as possible. In fact,
as we will see in Section 4, the model has some very interesting
properties at finite radius.

In Section 2 I will describe some techniques to calculate arbitrary
correlation functions
in the presence of a world-sheet cosmological constant (i.e.
time-independent tachyon background). To do this,
the algebraic structure of the model is discussed. I will also recall
the prescription introduced in \cite{klow} for finite radius
correlation functions and explain why it works
also in the deformed model.

In Section 3 some sample calculations are made to illustrate the
methods of Section 2. A $\langle PTTT\rangle$ correlation
function is calculated to genus 2 and an arbitrary tachyon
$n$-point function is calculated to genus 1.

In Section 4 I will study the partition function and the issue
of $R \rightarrow 1/R$ duality. A new kind of non-perturbative
duality is discovered. This should prove useful for a deeper
understanding of the theory.

\section{HOW TO CALCULATE CORRELATION FUNCTIONS}

\subsection{Ingredient One: The Algebraic Structure}

In \cite{jev} the matrix eigenvalue potential
$-\frac{1}{2\alpha '} x^{2}$ of the standard $c=1$ model is deformed
into $-\frac{1}{2\alpha '} x^{2} +\frac{\eta}{x^{2}}$.
Using the notation
in \cite{grop}, this gives the following oscillator algebra
for $\alpha ' =1$.
The operators are
\be
b=\frac{1}{2}(ip+x)^{2} -\frac{\eta}{x^{2}},
\ee
its conjugate $b^{\dagger}$ and the Hamiltonian
\be
H= \frac{1}{2}(p^{2}+x^{2})+\frac{\eta}{x^{2}}   .
\ee
They obey
$$
\left[ b, b^{\dagger} \right] =4H      ,
$$
\be
\left[ H,b\right]=-2b  \;\;\; \mbox{and}
\;\;\; \left[ H,b^{\dagger} \right]
=2b^{\dagger}                    .
\ee
I have, for convenience, continued
$\alpha ' \rightarrow -\alpha '$.
A special tachyon of momentum $p=2k$ is represented by $b^k$ (or
$b^{\dagger k}$ depending on chirality). The use of the corresponding
algebra in the undeformed case has been discussed
in [8--10].
It is nothing but the matrix-model $W_{\infty}$ algebra.
Through perturbation theory it is shown that
\be
\langle PT_{1}...T_{n} \rangle \sim
\langle P\left[ T_{1},\left[ T_{2},
...\left[
 T_{n-1} ,T_{n} \right] ... \right] \right] \rangle  \label{corr}
\ee
up to the factorized external leg factors;  $T_{1}$ through $T_{n-1}$
has positive chirality while $T_{n}$ has negative.
The right-hand side two-point function is given by
\be
\langle PW\rangle =-\frac{1}{\pi} {\cal I} \sum_{n=0}^{\infty}
\frac{\langle n \| W \| n \rangle}{E_{n}-\mu};\label{pw}
\ee
$n$ labels the one-particle states in the potential.
This also applies to
the deformed model. To obtain an $n$-point function,
we hence need the commutator
\be
\left[ b^{\dagger k_{1}},\left[ b^{\dagger k_{2}},
...\left[b ^{\dagger k_{n-1}}, b^{l}
\right] ... \right] \right]       ,
\ee
with $\sum _{i=1}^{n-1} k_{i} =l$.

The following important relations can be obtained after some work
\be
\left[ b^{l},b^{\dagger k} \right] = \sum _{r=0}^{k-1} 4^{k-r}
\left( \begin{array}{c} k\\r \end{array} \right)
(l)_{r-k} b^{\dagger r} b^{l-k+r} (H-l+k)_{k-r}  \label{komm1}
\ee
and
\be
f(H)b^{\dagger k} =b^{\dagger k} f(H+2k)    \label{komm2}
\ee
for any function $f(H)$. I use the notation
$(x)_{k}=x(x+1)...(x+k-1)$ and
$(x)_{-k} =x(x-1)...(x-k+1)$ for $k$ positive.

These are all the commutation relations that we need.

\subsection{Ingredient Two: The Matrix Elements}

To complete
the calculation we also need some matrix elements.
As shown in
\cite{grop}, we have
\be
\langle n \| b^{k} b^{\dagger k} \| n \rangle =
2^{2k}\left( \frac{n}{2} +a +1\right) _{k}
\left( \frac{n+1}{2}\right) _{k}
\ee
and
\be
\langle n \| b^{\dagger k} b^{k}  \| n \rangle =
2^{2k}\left(
\frac{n}{2} +a +1-k\right)
_{k} \left( \frac{n+1}{2}-k\right) _{k}  .
\ee
Here $a= -\frac{1}{2} + \sqrt{\frac{1}{4}+2\eta }$.
These matrix elements will appear as part of a two-point function,
as in (\ref{pw}). As explained in \cite{grop,dero}, the sum over
states in (\ref{pw}) should only be over {\it odd} $n$. This
is to make sure that all wave functions are normalizable.
The sum is obtained by using the formal equality
\be
\sum_{n=0}^{\infty} \frac{f(n)}{n+z}  = f(-z) \sum _{n=0}^{\infty}
\frac{1}{n+z}                                    .
\ee
The sum is defined by taking derivatives in $z$.
This gives $n \rightarrow -y+i\mu$,
where $y=\sqrt{\frac{1}{4}+2\eta}$, since,
as explained in \cite{grop},
$E_{n} =\frac{1}{2i\sqrt{\alpha '}}(2n+1+2a)$ in the deformed model.
The imaginary energies
arise from the convenient continuation $\alpha ' \rightarrow
-\alpha '$ introduced in \cite{nick}.
Powers of $H$ in the matrix elements are easily taken care of
inside the sum according to
\be
\langle n \| b^{\dagger k} b^{k} H^{l} \| n \rangle \rightarrow
(i\mu ) ^{l} \langle n \| b^{\dagger k} b^{k} \| n \rangle  .
\ee
If we expand in large $y$
(corresponding to large $\eta$ and hence weak coupling) we
find


\bea
\lefteqn{
\sum _{n odd} ^{\infty}
\frac{\langle n \| b^{\dagger k}b^{k} H^{l}
\| n \rangle }
{E_{n} -\mu} =}       \nonumber \\
&
(-1)^{k} y^{2k-4}(i\mu ) ^{l} \left[
y^{4} -
\left(
\frac{4k^{3}-k}{3} -2k^{2}i\mu
-k\mu ^{2} \right)
y^{2} \right.  \nonumber   \\
&
+\frac{k(k-1)(2k-1)(2k-3)(2k+1)(10k+7)}{90} +
\frac{2k^{2}(k-1)(3+2k-4k^{2})}{3}i\mu    \nonumber \\
&
\left.
-\frac{k(k-1)(10k^{2}-2k+3)}{3} \mu ^{2} +2k^{2}(k-1)i\mu ^{3}+
\frac{k(k-1)}{2} \mu ^{4}+...\right]   \nonumber \\
&
\times \sum _{n odd}^{\infty} \frac{1}{E_{n}-\mu}       .
\eea
This will be enough for all calculations to genus 2.

\subsection{How to Obtain Finite-Radius Results}

In this section I will discuss how to obtain correlation functions
at finite radius. In \cite{klow} it was shown that the
finite-radius correlation functions can be obtained from the
non-compactified ones by acting with
\be
\frac{\frac{1}{R}\frac{\partial}{\partial \mu }}
{e^{\frac{i}{2R}\frac{\partial}{\partial \mu }}-
e^{-\frac{i}{2R}\frac{\partial}{\partial \mu }}}
\sim 1+\frac{1}{24R^{2}} \frac{\partial ^{2}}{\partial \mu ^{2}} +
\frac{7}{5760R^{4}} \frac{\partial ^{4}}{\partial \mu ^{4}}
+...
\label{rad}
\ee
The easiest way to see this, \cite{avhand}, is to consider the
puncture two point function $<PP>$ given by
\be
\frac{1}{\pi R} {\cal R}
\sum _{n,m=0} ^{\infty} \frac{1}{(\frac{2n+1}{2\sqrt{\alpha '}}
+\frac{2m+1}{2R} +i\mu )^{2}}  \label{part}     .
\ee
This can be obtained from the non-relativistic fermionic field theory
description of the $c=1$ matrix model \cite{ferm}, and an equivalent
expression was first derived in \cite{grokle}.
The $R \rightarrow \infty$ limit of this is clearly
\be
-\frac{1}{\pi} {\cal R}
\sum _{n=0} ^{\infty} \frac{1}{\frac{2n+1}{2\sqrt{\alpha '}}
+i\mu}      .                         \label{pjutt}
\ee
Applying  (\ref{rad}) on (\ref{pjutt}) gives (\ref{part}).
Other correlation functions are obtained through perturbation theory,
a procedure that clearly commutes with (\ref{rad}).
This was shown in the standard $c=1$ case. However, as should be
clear from above, the same reasoning applies to the case of the
deformed
matrix model. Just recall that
\be
\langle PP \rangle =   \frac{1}{\pi R} {\cal R}
\sum _{n,m=0} ^{\infty} \frac{1}{(\frac{2(2n+1)+1+a}{2\sqrt{\alpha '}}
+\frac{2m+1}{2R} +i\mu )^{2}}      \label{purt}
\ee
and it is clear that everything goes through.

\section{SOME EXAMPLES}

\subsection{Four-point Correlation Function to Genus 2}

In this section I will calculate some correlation functions
at finite radius. I will begin with
$\langle PTTT \rangle $ up to genus 2.
That is, the tachyon four-point function
with one of the momenta put to zero.

The needed commutator is
\be
\left[ b^{\dagger k_{1}}, \left[ b^{\dagger k_{2}} , b^{l}
\right] \right] =
b^{\dagger k_{1}+k_{2}}b^{l} -
b^{\dagger k_{1}} b^{l} b^{\dagger k_{2}} -
b^{\dagger k_{2}} b^{l} b^{\dagger k_{1}} +
b^{l} b^{\dagger k_{1}+k_{2}}     ,
\ee
where $l=k_{1}+k_{2}$. Using (\ref{komm1}) and (\ref{komm2})
we find
$$
\sum _{s=0}^{l-1} 4^{l-s}
\left( \begin{array}{c} l\\s \end{array} \right)
(l) _{-l+s} b^{\dagger s}b^{s} (H)_{l-s}
-\sum _{s=0}^{k_{1}-1} 4^{k_{1}-s}
\left( \begin{array}{c} k_{1}\\s \end{array} \right)
(l) _{-k_{1}+s} b^{\dagger k_{2}+s}b^{k_{2}+s}
(H-k_{2})_{k_{1}-s}
$$
\be
-\sum _{s=0}^{k_{2}-1} 4^{k_{2}-s}
\left( \begin{array}{c} k_{2}\\s \end{array} \right)
(l) _{-k_{2}+s} b^{\dagger k_{1}+s}b^{k_{1}+s}
(H-k_{1})_{k_{2}-s}      .
\ee
One must then use Section 2.2.
The result must also be reexpanded in terms of $\eta$ instead of $y$.
Furthermore, we should recall that $p=2k$ at $\alpha ' =1$.
When all this is done, and the normalization fixed to that
of collective field theory, we find
\bea
\lefteqn{\langle PTTT \rangle \sim } \nonumber \\
&
pp_{1}p_{2} \left(
\eta ^{p/2-1}-
\frac{1}{48} (p-2)(p^{2} +3(p_{1}^{2}+p_{2}^{2}) -4p -15)
\eta  ^{p/2-2}
+\frac{3}{4}(p-2)\mu ^{2}\eta ^{p/2-2}   \right. \nonumber
\\ &
+\frac{(p-2)(p-4)}{23040} (80(p_{1}^{4}+p_{2}^{4})-288
(p_{1}^{3}+p_{2}^{3})-728(p_{1}^{2}+p_{2}^{2})+1176p+2085
\nonumber \\ &
+240p_{1}^{2}p_{2}^{2}+80(p_{1}p_{2}^{3}+p_{1}^{3}p_{2})
+384(p_{1}p_{2}^{2}+p_{1}^{2}p_{2})+44p_{1}p_{2} )  \eta ^{p/2-3}
\nonumber \\ &
-\frac{(p-2)(p-4)}{64}
(l^{2}+5(p_{1}^{2}+p_{2}^{2})-8p-23)\eta ^{p/2-3} \mu ^{2}
\nonumber \\ & \left.
+\frac{5}{32}(p-2)(p-4)\eta ^{p/2-3} \mu ^{4} +... \right)  .
\eea
Here a zero due to the single puncture has been extracted.
A comment about the genus expansion is needed. It is not only
$\eta \sim 1/g^{2}$ that carries powers of the string coupling
constant, we also have $\mu \sim 1/g$. This means, for instance,
that the $\mu ^{2} \eta ^{p/2-2}$ term above is at genus 0.
Indeed, taking two $\mu$ derivatives gives $\langle PPPTTT \rangle$
at genus 0. Hence, from this point of view, the above expression,
which only keeps the first few terms, is not complete at a fixed genus.
However, it includes everything that is needed when we turn to
finite radius at zero $\mu$.

As we saw in Section 2.3.
the finite radius expression at $\mu =0$ is obtained immediately
by substituting
\be
\mu ^{2} \rightarrow \frac{1}{12R^{2}}
\ee
and
\be
\mu ^{4} \rightarrow \frac{7}{240R^{4}}      .
\ee
This completes the calculation of $\langle PTTT\rangle$ to genus 2
at finite radius.

\subsection{$n$-point Correlation Function to Genus 1}

Next let me consider a general $n$-point function. This time to genus 1.
However, I will begin with $\langle PT_{1}...T_{n}\rangle $
where $n$ is even. This
will allow us to verify, as noted in \cite{jev,dero}, that odd-point
functions vanish when $\mu =0$.

Generalizing the above calculation, one finds
\be
\langle PT_{1}...T_{n}\rangle  \sim
\sum _{i=1}^{n-1} A(k_{i}) -
\sum _{i_{1}>i_{2}}^{n-1} A(k_{i_{1}}+k_{i_{2}}) +
\sum _{i_{1}>i_{2}>i_{3}}^{n-1} A(k_{i_{1}}+k_{i_{2}}+k_{i_{3}}) -...
\ee
where
\be
A(k) = \sum _{s=0}^{k-1} 4^{k-s} \left( \begin{array}{c}
k\\s \end{array} \right)
(l)_{-k+s} b^{\dagger l-k+s} b^{l-k+s} (H-l+k)_{k-s}   .
\ee
As before $\sum _{i=1}^{n-1} k_{i} =l$.
If we are interested only in lower genera, we should keep only the
terms with the highest values of $s$. We can then write

\bea
\lefteqn{
\langle PT_{1}...T_{n}\rangle  \sim } \nonumber \\
&
\frac{4^{\frac{n}{2}}}{(\frac{n}{2})!}
(l)_{-\frac{n}{2}}
B_{n,0} b^{\dagger l-\frac{n}{2}} b^{l-\frac{n}{2}}
+\frac{4^{\frac{n+2}{2}}}{(\frac{n+2}{2})!}
(l)_{-\frac{n+2}{2}}
B_{n,1} b^{\dagger l-\frac{n+2}{2}} b^{l-\frac{n+2}{2}}
+...
\eea
where
\bea
\lefteqn{B_{n,m}=
\sum _{i=1}^{n-1}
(k_{i})_{-\frac{n}{2}-m}(i\mu -l+k_{i})_{\frac{n}{2}+m}  }
\nonumber \\ &
 -\sum _{i_{1}>i_{2}}^{n-1}
(k_{i_{1}}+k_{i_{2}})_{-\frac{n}{2}-m}
(i\mu -l+k_{i_{1}}+k_{i_{2}})_{\frac{n}{2}+m}
+...
\eea
Terms with higher powers of $b^{\dagger}$ and $b$ vanish.
One can calculate that
\be
B_{n,0} = \frac{n!}{2} i\mu k_{1}...k_{n-1}
\ee
and
\bea
\lefteqn{B_{n,1} =
\frac{n!(n+2)}{24} i\mu k_{1}...k_{n-1}    }
\nonumber \\ &
\times \left[
-\frac{n^2+n}{2} +\frac{3n}{2} \sum _{i=1}^{n-1} k_{i}+
\frac{n-2}{2} \sum _{i=1}^{n-1} k_{i}^{2}
\right.
\nonumber \\ &
\left.
-3\sum _{i>j}^{n-1} k_{i}k_{j}+
3i\mu \left(
\sum _{i=1}^{n-1} k_{i} -\frac{n}{2}\right)
-\frac{n-2}{2} \mu ^{2} \right]
{}.
\eea
It is seen that the correlation function will be proportional
to $\mu$ and vanish at $\mu =0$ unless we take a $\mu$ derivative,
i.e. insert an extra puncture.
Using the results obtained in this way for
$\langle PPT_{1}...T_{n}\rangle$  it is easy to generalize to
$\langle T_{1}...T_{n}\rangle$.
The final answer is, after some work,
\bea
\lefteqn{
\langle T_{1}...T_{n} \rangle \sim
(n-3)!! p(p-2)...(p-(n-4))\prod _{i=1}^{n-1} p_{i}
\left[ \eta ^{p/2 -n/2 +1}  \right.  }  \nonumber \\
&
\left.
-(p-(n-2))\left(
\frac{p^{2}+(n-1)\sum _{i=1}^{n-1} p_{i}^{2}
-2(n-2)p -4n+1}{48} -\frac{(n-1)}{48R^{2}}\right)
\eta ^{p/2-n/2} +... \right]
\eea
when normalized to collective field theory.
So, for the 2-point function
we find
\be
p\eta ^{p/2}
 -\frac{1}{48}p^{2}\left(
 2p^{2}-7-\frac{1}{R^{2}}\right) \eta ^{p/2-1} +...
\ee
and for the 4-point one
$$
pp_{1}p_{2}p_{3}\left( \eta ^{p/2-1}
 -\frac{1}{48}(p-2)\left(
 p^{2}+3(p_{1}^{2}+p_{2}^{2}+p_{3}^{2})
-4p -15 -\frac{3}{R^{2}}\right)
\eta ^{p/2-2}\right)
$$
\be
+...
\ee

\section{A NEW DUALITY SYMMETRY}

Duality is a property of the string theory partition function
(and puncture correlation functions). As observed in \cite{grokle}
the function
(\ref{part}) is invariant under
\be
\left\{ \begin{array}{c}
R \rightarrow \alpha ' /R   \\
\mu \rightarrow \frac{R}{\sqrt{\alpha '}}\mu
\end{array} \right.
\label{transf}
\ee
(i.e. $g \rightarrow
\frac{\sqrt{\alpha '}}{R} g$, where $g$ is the string coupling).
Correlation functions with non-zero momentum tachyons are clearly
not invariant, as discussed in \cite{klow}. Instead they are transformed
into correlation functions with non-zero winding.

Does the deformed matrix model also have a duality symmetry?
In \cite{dero} the answer was negative. Indeed, although the standard
duality transformation above (with $\mu ^{2}$ replaced by $\eta$)
works to genus 1 (for the partition function), it breaks down
for higher genera as shown in \cite{dero}. However,
I will show below the existence of a {\it new}
kind of non-perturbative duality. This is the most important result
in the paper.

Consider (\ref{purt}). By the same reasoning as for the undeformed case,
one finds a symmetry under
\be
\left\{ \begin{array}{c}
R \rightarrow \alpha ' /4R  \\
\frac{1+2a}{2\sqrt{\alpha '}}\rightarrow \frac{2R}{\sqrt{\alpha '}}
\frac{1+2a}{2\sqrt{\alpha '}}
\end{array}   \right.         ,
\ee
for the partition function and all puncture correlation functions.
Note that the self-dual radius is $R=\frac{1}{2} \sqrt{\alpha '}$
for this transformation,
rather than $R=\sqrt{\alpha '}$ as for (\ref{transf}).
This is due to the summation being done only
over odd states. The peculiar nature of this duality becomes
clear when we express it in terms of
$\eta \sim 1/g^{2}$.
We find
that the second transformation is
\be
\eta \rightarrow \frac{4R^{2}}{\alpha '} \eta +\frac{R^{2}}{2\alpha '}
-\frac{1}{8}                 .
\ee
Note the translational piece. Its presence implies
that the duality symmetry mixes different genera! A
partition function at a fixed genus is {\it not} invariant.
Only the full sum over all genera is. It is a simple exercise to check
the symmetry in the perturbative expansion. It is then seen how
contributions from different genera cancel against each other.

\section{CONCLUSIONS}

The duality symmetry discovered in this paper is difficult to
understand from a world sheet point of view. Duality is usually
thought to work genus by genus. The deformed matrix model tells us
that this might not always be the case. It will be very interesting to
understand the physical explanation, and
to find out what kind of string
theory have this peculiar property, regardless of whether it is
a black hole or not.

\section*{Acknowledgements}

I would like to thank  L. Alvarez-Gaum\'e and G. Veneziano for
discussions.

\end{document}